\documentclass[oneside]{book}%remove empty pages

\usepackage[authoryear]{natbib}
\usepackage[doublespacing]{setspace}

\usepackage{graphicx} 

\makeatletter
\newcommand{\chapterauthor}[1]{%
  {\parindent0pt\vspace*{-25pt}%
  \linespread{1.1}\large\scshape#1%
  \par\nobreak\vspace*{35pt}}
  \@afterheading%
}
\makeatother
\def\cite#1{\citep{#1}}

\begin{document}
\tableofcontents

\chapter[Fluorescent Silicon Clusters]{Fluorescent Silicon Clusters and
  Nanoparticles}
\chapterauthor{Klaus von Haeften}
\section[Overview]{Overview}

Clusters, consisting of a small number of atoms, have been in the focus of
physical and chemical research for several decades. They often show
dramatic size effects. The 
addition of a single atoms can change their properties rather
abruptly because of, for example, the discreteness of shell filling
\cite{knight1984} or sphere-packing effects \cite{echt1981}. 
When clusters become larger and reach the nanometre scale, other
effects are observed, such as quantum confinement; 
The intense red fluorescence observed for nanostructured silicon
\cite{canham1990,cullis1991,wilson1993,lockwood1994,cullis1997} is a popular and
frequently cited example of this effect. The discovery of
fluorescent nanoscale silicon at room temperature by Canham
\cite{canham1990} increased the already quite intense research into silicon
clusters further, and to date numerous examples of nanostructured forms 
of fluorescent silicon have been reported
\cite{takagi1990,brus1995,hirschman1996,borsella1997,ehbrecht1997,cullis1997,huisken1999,pavesi2000nature,belomoin2000,belomoin2002,mangolini2005,falconieri2005,brewer2009,vincent2010,he2011,dasog2014,li2016}. Hence, we have a rich set of data available on electronic and structural
properties that unpin our understanding of the fluorescence of silicon
clusters and nanoparticles.

The strong appeal of fluorescent silicon clusters and nanoparticles arises
due to a veritable multitude of applications, for example in electronic circuits
\cite{pavesi2003rev,svrcek2004,stupca2007}
%, optical fibre communication technologies
%\cite{kenyon2005,bakker2015,dmitriev2016}%not fluorescent 
and biomedicine \cite{nel2009,chinnathambi2014}. Silicon is the most
frequently used semiconductor material in
electronics. It has been suggested that by combining electric and
optical signal transmission, higher performance can be achieved
\cite{canham2000,pavesi2003rev}. Issues directly related to the decreasing size of
electronic units, such as signal delay caused by increasingly longer
interconnects, can be addressed by optically transmitted refresh pulses 
\cite{pavesi2003rev}. Fluorescent 
silicon clusters and nanoparticles are also attractive in biomedical
applications because nanoscale silicon and silicon dioxide are considered to be
non-toxic, or at least considerably less harmful than fluorescent
nanoparticles of other materials
\cite{warner2005,erogbogbo2007,erogbogbo2008,choi2009,erogbogbo2010,cheng2014,peng2014,mcvey2014}
as well as biodegradable \cite{park2009}. Fluorescent
silicon nanoparticles play an important role as biological markers
\cite{nel2009,montalti2015}.  

%The findings from research into porous silicon fluorescence revealed
%that, in broadest terms, the ability of silicon to fluoresce is
%governed by the following principles: a
%sufficiently small size, changing the bulk band structure from indirect
%to quasi direct, and specific surface properties, prohibiting
%non-radiative decay to the ground state which is in competition to
%fluorescence. 

In this chapter, the foundations of silicon cluster experiments and
cluster production will be discussed. The underlying principles behind the
fluorescence of silicon clusters and nanoparticles, such as quantum
confinement and surface passivation, are introduced, and contrasted with
the current state of research on fluorescent silicon 
clusters and nanoparticles. Owing to the vast number of
publications on this subject that can already be found in the
literature, this book chapter cannot be exhaustive. Rather, it 
complements recent review articles on fluorescent nanoscale silicon
\cite{dohnalova2014,cheng2014,priolo2014,peng2014,mcvey2014,montalti2015,dasog2016}.

\section[Fundamental concepts]{Fundamental concepts for
  producing fluorescent nanoscale silicon}\label{sec:concepts}

\begin{figure}%[!ht] 
\centering
\includegraphics[width=0.9\columnwidth]{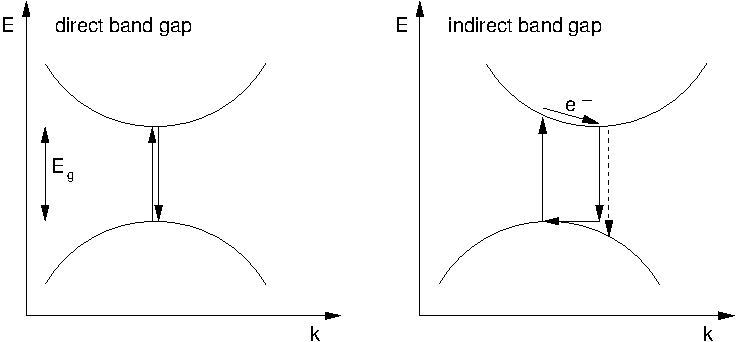}
\caption{
Schematic of fluorescence emission in direct and indirect band gap
semiconductors. The scheme shows the valence (bottom) and conduction bands
(top) of a direct (left) and an indirect band gap semiconductor
(right). The arrows indicate the magnitude of the energy band
gap, E$_g$, the pathways of excitation of an electron from the valence
band to the conduction band, electronic relaxation, fluorescence and
absorption of phonons (see text).
} 
\label{fig:direct-indirect}
\end{figure}

Bulk crystalline silicon is known as a poor light emitter because of its
indirect band gap. Fluorescence is a relatively inefficient relaxation process in
electronically excited indirect band gap semiconductors because
fluorescence has to simultaneously occur with the absorption of a phonon
of matching momentum. This mechanism is illustrated in more detail in
figure~\ref{fig:direct-indirect}. The entire electronic excitation and
relaxation/fluorescence cycle is shown for direct and indirect band
gap semiconductors.

The left-hand side of figure~\ref{fig:direct-indirect} shows an energy
band schematic of a direct band
gap semiconductor, characterised by the conduction and valence band maxima
and minima being on top of each other. Photoexcitation of an electron follows
an energetic pathway indicated by the vertical arrow, reaching from the top of the valence band an into the conduction
band, from where it returns to recombine with the hole, inducing
fluorescence. Photoexcitation using higher energies is possible, but
less likely because of the decreasing density of states along the
parabola, and indeed the fluorescence wavelength will remain unaltered because of
the relatively short timescale on which electronic relaxation occurs in the conduction and valence bands. The
fluorescence intensity will be unchanged because both electron and hole have
the same momentum.

This situation changes in indirect band gap semiconductors, as shown on
the right hand side in figure~\ref{fig:direct-indirect}. In indirect
band gap semiconductors, the minima of the conduction band and maxima of the valence
band are shifted. As a consequence, electrons that are
photoexcited into the conduction bands quickly undergo electronic
relaxation to the minimum of the conduction band. However, any subsequent direct, vertical decay
to the valence band is not possible because all states with similar
momenta, $\hbar k$, are populated with electrons; in other words, hole
states for direct recombination with the excited electrons are not
available. 'Diagonal' recombination with the available, original hole
state at the maximum of the  valence band is not an option because momentum would no longer be conserved.
So, in order for diagonal recombination to 
happen, a phonon with matching momentum has to be absorbed during
fluorescence, but because these two processes would have to happen
simultaneously, this scenario is clearly a rare occurrence. As a consequence, the
fluorescence lifetime from bulk crystalline silicon is very long and
fluorescence intensity is very weak.  

When bulk crystalline silicon is reduced in size to a scale approaching that of the
nanoscale regime, translational symmetry is gradually lost. A
consequence of small size of nanocrystals is, therefore, that materials
that are indirect band gap semiconductors in the bulk phase become
quasi-direct semiconductors at the nanoscale. For silicon, this means
that high fluorescence intensities are possible if nanocrystallites are only
small enough to display a quasi-direct band gap.

Another feature of such reductions in crystal size is quantum
confinement. Quantum confinement relates to the shift of energy levels
with size, and neatly explains the energy spectrum of, for example,
colour-centre defects in crystals \cite{hayes2004,fox2010}, and electrons
confined in nanoscale bubbles in liquid helium
\cite{fowler1968,grimes1990,grimes1992}. 
 
These energy shifts can be understood using the popular
particle-in-a-box model that is almost perennially reviewed in quantum mechanical text books. The
energy spectrum, $E(n)$, of an electron confined in a one-dimensional
box of length $\ell$ with infinitely high box walls can be
straightforwardly derived to give the following equation:  

\begin{equation}
E(n) = \frac{n^2 \pi^2 \hbar^2}{2 m_e \ell^2} \label{eq:box}
\end{equation}
where $n$ is the principal quantum number, $\hbar$ the reduced Planck (or Dirac)
constant and $m_e$ the electron mass. The quantum number, $n$, is indexed from
$n$~=~1, and the energy difference E($n$~=~2) and E($n$~=~1) would be
equivalent to the fluorescence energy from the first excited state to
the ground state.

The analogy of the one-dimensional model can be straightforwardly extended to three dimensions. A more realistic model,
using the work function rather than infinitely high potential walls,
requires solving transcendental equations. This latter, more rigorous, treatment
lowers the energy values by not more than 10~\%, for which decreases in the energy difference and
effective magnitude of the box
diameter for a given transition energy are implicit. Furthermore, a more realistic potential surface than a square well would also lower the energy levels as the constraints imposed by the former must, of their very nature, be more relaxed. 

Lockwood and co-workers observed that the fluorescence of a Si/SiO$_2$
superlattice depended on the silicon film thickness
\cite{lockwood1996}. They attributed this behaviour to one-dimensional
confinement of the excited electron within the silicon film. To explain
the shift in the fluorescence energy, they adopted the particle-in-a-box
model and showed that the peak energy of the observed red
fluorescence band followed equation~\ref{eq:lockwood}.
\begin{equation}
E(n) = E_g + \frac{\pi^2 \hbar^2 }{2 d^2} \left(\frac{1}{m_e^*} +
\frac{1}{m_h^*}\right) \label{eq:lockwood}
\end{equation}
Here, $d$ is the silicon layer thickness, $m_e^*$ and $m_h^*$ are the
'effective masses' of electron and hole, although the authors
acknowledge that, strictly speaking, the concept of effective electron
and hole masses has 
no physical meaning in nanoscale systems that do not exhibit the translational
symmetry of crystals. A good fit was reported for $E(eV) = 1.60 + 0.72
d^{-2}$~eV, with $d$ given in nm, which implies effective masses
$m_e^* \approx m_h^* \approx $~1~m$_e$ 
in good agreement with bulk crystalline silicon were
$m_e^*(Si_{bulk})$~=~1.18~m$_e$ and $m_h^*(Si_{bulk})$~=~0.81~m$_e$. The
fit shows that the band gap energy, at $E_g$~=~1.60~eV, is considerably larger than that
of bulk crystalline silicon ($E_g$(c-Si)~=~1.12~eV at 295~K), and is, in fact, more
similar to that of amorphous silicon ($E_g$(a-Si)~=~1.5 - 1.6~eV at
295~K). The good fit with experimental data and the similarity between
the confinement term in equation~\ref{eq:lockwood} and the original
particle-in-a-box equation~\ref{eq:box} is remarkable, and strongly
supports the presence of quantum confinement.

Park and co-workers investigated amorphous silicon quantum dots embedded
in silicon nitride. They observed fluorescence in the form of a single
band whose maximum shifted with average quantum dot size. The
size dependence of fluorescence energy fit to the equation was found as $E(eV) = 1.56 + 2.40
d^{-2}$~eV, which confirmed the earlier observed band gap energy of
amorphous silicon \cite{lockwood1996} in the limit of large dot
sizes. However, the dependence on quantum confinement, $2.40
d^{-2}$~eV, was much larger. The discrepancy with work on conducted on Si/SiO$_2$
superlattices was attributed to the three-dimensional confinement of the
quantum dots in silicon nitride \cite{park2001}.

Summarising, we have so far seen that with reduction of size of a
crystal translational symmetry is gradually lost with the consequence
that indirect band gap semiconductors become quasi-direct semiconductors
at the nanoscale. At the same time the band gap energy increases because
of quantum confinement.

A further important factor determining the ability to fluoresce is the
electronic structure at the surface of nanocrystals. Surfaces break
translational symmetry, a consequence of which is that one cannot expect the
same energy band structure as might be observed for 'infinite' crystals. Band gap energies may
be smaller, or may not even exist. For nanocrystals, this means that
surface effects may compete with quantum confinement. In the following,
non-radiative decay at nanocrystal surfaces will be discussed.

\begin{figure}%[!ht] 
\centering
\includegraphics[width=0.9\columnwidth]{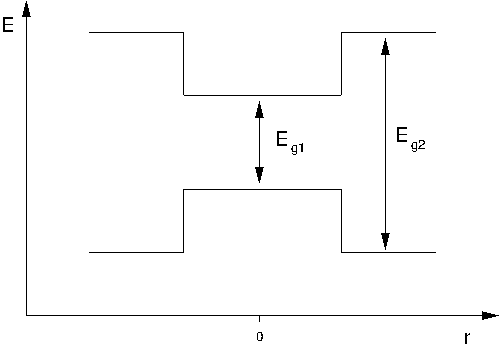}
\caption{
Schematic of a nanoscale crystal (quantum dot) embedded in a material
of larger band gap energy. E$_{g1}$ and E$_{g2}$ refer to the band gap
energies of the nanocrystal and host material, respectively.
} 
\label{fig:bg}
\end{figure}

To prevent non-radiative decay at its surface, a nanocrystal may be
embedded in another semiconductor or insulator of larger band gap
energy. This effect is illustrated in figure~\ref{fig:bg}.

In broadest terms, one can expect that surface effects
on the band gap energy can be minimised for such a system . It is assumed that the
nanocrystal structure fits well with that of the host and that a minimum
of additional interface states are produced. Ideally, this would mean
that an electron promoted across E$_{g1}$ from the valence band of the
nanocrystal to its conduction band would remain confined within the
nanocrystal. The electron would have no other choice than to fluoresce
to the ground state because no discrete states are available within the gap. 

\begin{figure}%[!ht] 
\centering
\includegraphics[width=0.9\columnwidth]{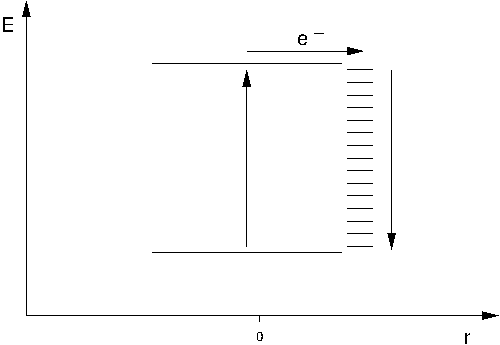}
\caption{
Simplified energy level diagram of a nanoscale crystal with a high density of
vibrational states at its surface. The small size of the crystal means
that electrons delocalised over the conduction band quickly localise at
the surface. They then relax non-radiatively by hopping across the
vibrational states. 
} 
\label{fig:relax}
\end{figure}

To better account for non-radiative decay in real systems, vibrational
relaxation is often considered. Because of the possibility of surface
reconstruction, vibrational relaxation at surfaces is particularly
important for free nanoscale crystals and
clusters. Figure~\ref{fig:relax} illustrates the cycle of excitation,
electronic migration and relaxation within the conduction band and
vibrational relaxation at the surface. In this simple picture, a high
density of vibrational states at the surface is assumed to exist. The
electronically excited electron 'finds' the surface on a
sub-femtosecond timescale, and relaxes by jumping down the energy ladder
of vibrational states.

Another simplification implicit in this picture of a dense spectrum of
vibrational states is that the short timescales of non-radiative decay
are ignored. The energy level scheme shown in figure~\ref{fig:relax}
implies that the vibrational levels are time-averages.

\begin{figure}%[!ht] 
\centering
\includegraphics[width=0.9\columnwidth]{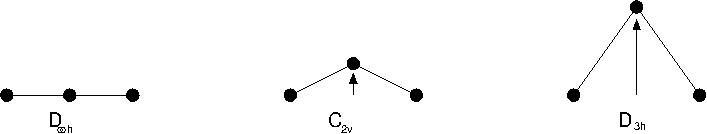}
\caption{
Simple schematic illustrating how vibrational-electronic coupling can lead to
non-adiabatic relaxation from electronically excited states to the
ground state. A linear, homonuclear triatomic molecule performing a
bending vibration assumes, at various points during one such oscillation, structures that belong to three different
point groups, and hence three different electronic states. 
} 
\label{fig:symmetry}
\end{figure}

Under closer inspection, the
consideration of relaxation pathways along such eigenstates in
nanocrystals does not appear to be justified. Particularly in free clusters, the
atoms at the surface can perform large amplitude vibrations. These large
amplitude vibrations give rise to electronic-vibrational coupling, and
open other, non-adiabatic routes for relaxation from electronically excited states to the
ground state. The simple example of the triatomic, homonuclear molecule
shown in figure~\ref{fig:symmetry} illustrates this mechanism.
  
In this example, the molecule performs a bend vibration. At the
classical turning points of this motion, structures belonging to three different point
groups, D$_{\inf h}$, C$_{2v}$ and D$_{3h}$, can be identified, with the actual point
group depending on the amplitude of the vibration. In other words, by
performing a bend vibration, the molecule is able to intersect the different
electronic states associated with these point groups. These
intersection points provide passages to lower-lying states. 

These lower states may be dissociative. Therefore, a tendency towards
isomerisation and dissociation may be expected, particularly for small
homonuclear clusters. This corroborates the relevance of caging through
a rigid shell of atoms, moieties or solvent molecules. In fact, deposition of
metal clusters into argon matrices and argon droplets has made the
observation of photoluminescence possible
\cite{felix2001,sieber2004,conus2006,harbich2007}. Argon shells 
have also been used to cage oxygen molecules; the observed luminescence was
attributed to oxygen atoms that had first dissociated but were then
forced, by the cage, to recombine \cite{laarmann2008cage}.

In summary, non-adiabatic decay at the surface of homonuclear clusters
is an important relaxation channel in terms of its competition with fluorescence. The
design and engineering of fluorescent clusters therefore requires that this possibility be suppressed. 
In broadest terms, the passivation of dangling
bonds at a surface with atoms such as hydrogen (hydrides), halogens
(halides), oxygen (oxides), or molecular groups such as alkenes, amines
or polymers, can be understood to confine the electronically excited
electron to the core region of the cluster. The distinctly higher fluorescence
intensities that arise from this confinement can be observed. This was confirmed by Seraphin and coworkers
\cite{seraphin1996} who investigated silicon nanoparticles produced by
laser ablation with and without passivation in vacuo. They found that passivation increased fluorescence intensity,
but left the fluorescence energy itself unchanged.
Embedding clusters and
nanocrystals within a lattice-matching host of a larger band gap
energy fulfils the same purpose. Also, in the ideal case of a
lattice-matching interface, large amplitude vibrations would be
successfully suppressed purely as, though not necessarily limited to, a matter of sterics. Passivated or core-shell clusters embedded in a
solvent can also effectively transfer and absorb excess vibrational energy.

\section[Clusters]{Clusters}\label{sec:cluster}
Clusters are understood as being particles consisting of as few as two, three, or four, or as having
as many as a few thousand, atoms. The term 'cluster'  is used alongside the term
'nanoparticles', but clusters are commonly understood to represent smaller
particles. The study of clusters is motivated by the 
desire to understand the often dramatic changes of material
properties that accompany changes in size. Material properties also depend
on structure and dimensionality, and similarly accompany changes in size. All such characteristics are relevant in cluster science research.  

\subsection[Production]{Production of silicon clusters}

Free clusters are frequently studied in supersonic beams. A gas under
high pressure is expanded through a tiny aperture into vacuum. During
expansion, it cools and nucleates via removal of excess energy through three-body collisions, forming
clusters. They fly through vacuum where they can, for instance, be investigated free
from external interactions, or can otherwise be deposited on a substrate. For silicon
clusters, this production method is unsuitable because it requires atomic
vapour of significant pressure. Silicon has a comparably low
vapour pressure, even at high temperatures, rendering supersonic
expansion of thermally produced silicon vapour through a nozzle into
vacuum an unrealistic mechanism for silicon cluster production.

Supersonic beams have been employed in sources where silicon has been
generated by decomposition of precursor gases which are then mixed with a
carrier/aggregation gas. Silane diluted in helium 
and exposed to a discharge has been been expanded through a pulsed
piezoelectric valve \cite{hoops2001}. Ehbrecht et al.\ used a pulsed
CO$_2$ laser to decompose silane diluted in helium
\cite{ehbrecht1999}. The products were 
expanded into vacuum, and subsequently investigated by time-of-flight mass spectrometry
or deposited onto CF$_2$ or LiF substrates \cite{ehbrecht1997}.

Silicon clusters have been produced using the principle of gas
aggregation \cite{sattler1980}. Silicon vapour is mixed with a 'seeding'
gas, which induces the three body collisions required for
nucleation. It also facilitates growth by mediating collisions between
silicon atoms and removing the latent heat that is released during
any subsequent growth of clusters from these collisions.

To generate silicon vapour within a seed gas, laser vaporisation,
sputtering and pulsed arcs have been used. Laser vaporisation sources
employ pulsed laser sources that are fired at a rotating silicon rod
\cite{bloomfield1985,jarrold1991prl,bower1992,honea1993zpd,honea1999}. 
Sputter gas aggregation sources employ a modified sputter 
electrode configuration, allowing higher operating gas pressures than
usually needed in sputtering for thin film production
\cite{haberland1991,astruchoffmann2001ups,pratontep2005,wegner2006,haeften2009}. 
Argon and helium are used as seed gases. Helium produces smaller clusters
because it is much lighter than silicon and energy transfer is less
efficient. The pulsed arc cluster ion source is another variation of a
gas aggregation-type cluster source that has been used for silicon
cluster production \cite{maus2000}.

These sources produce neutral and ionic clusters, including both cations and
anions; charged clusters can be easily manipulated in a spatial sense via electrostatics, and can therefore be mass-selected. Kitsopoulos
and co-workers size-selected silicon cluster anions, which were
then photoexcited into the neutral ground state as well as other low-lying
electronically excited states of the neutral clusters by the process of
electron detachment \cite{kitsopoulos1990}.
Honea and
co-workers used arrangements of linear and perpendicular quadrupole mass
selectors to size-select silicon cluster ions and deposit them onto a
surface \cite{honea1999}. To soften the impact, rare gas buffer layers
were co-deposited onto a cold, pre-deposited rare gas matrix \cite{honea1999}. A similar scheme
was used by Grass 
and co-workers, who soft-landed Si$_4$ on highly oriented pyrolytic
graphite (HOPG) and performed X-ray photoelectron spectroscopy
\cite{grass2002}.  
Astruc-Hoffmann and co-workers used a reflectron time-of-flight (RETOF) mass
spectrometer in combination with a multi-wire mass gate to size select
silicon anion clusters on which to subsequently perform photoelectron spectroscopy. 
The size-selected clusters were irradiated with UV light from an
ArF excimer laser and analysed in a magnetic bottle photoelectron
spectrometer \cite{astruchoffmann2001ups}. 

Hirsch and co-workers introduced size-selected silicon clusters into an ion trap
\cite{hirsch2009,lau2011,vogel2012,kasigkeit2015}. The trapped clusters
were excited by monochromatic synchrotron radiation, allowing the
determination of core-level shifts for specific cluster sizes.

\subsection[Geometric structure]{Geometric structure}

A great variety of geometric structures have been reported for silicon
clusters using both theoretical and experimental methods. in general,
it has been found that the structures of neutral, cation and anion cluster species differ considerably. Anions
have frequently been used to elucidate electronic and geometric
features. Their structures are affected by Jahn-Teller distortions.
Care must be taken when compared to neutral clusters.

% \cite{rinnen1992}.
%{rinnen1992} absorption of Si_n is size independent, photodissociation,
%conclusion: diamond like. Contradictory to most other work
In a number of early studies, silicon clusters produced using supersonic beam techniques were
deposited and investigated spectroscopically. Honea and coworkers
deposited size-selected silicon clusters into argon, krypton and nitrogen
matrices using co-deposition onto a liquid helium-cooled substrate
\cite{honea1993,honea1999}. Using surface-plasmon-polariton-enhanced Raman
spectroscopy, sharp features characteristic of Si$_4$, Si$_6$ and Si$_7$ 
 structures were identified in their spectra, which were in good agreement with
earlier ab initio calculations
\cite{raghavachari1985,raghavachari1986,tomanek1986,raghavachari1988,ballone1988}. The 
spectra also revealed evidence for the presence of cluster-cluster
aggregation within the rare gas matrix. The spectra of the aggregates of
Si$_4$, Si$_6$ and Si$_7$  bore considerable similarities to those of larger
clusters, such as Si$_{25-35}$, as well as those of amorphous silicon \cite{honea1999}.

The structure of free silicon clusters in the size range from n~=~10~-~100
was addressed using drift mobility measurements, the results of which revealed a
prevalence for prolate-shaped structures. A structural transition
occurs at sizes around n~=~27, however; larger clusters were found to prefer more
spherical configurations \cite{jarrold1991prl}. The preference for prolate
shapes was attributed to a tendency of the silicon atoms to minimise
coordination. This trend competes with the surface energy of the
clusters, ultimately leading to a preference for spherical structural
motifs in larger clusters \cite{jarrold1991prl}. The experiments were
repeated at higher resolution \cite{hudgins1999}, and calculations
confirm these experimental findings \cite{ho1998,jackson2004}.

Vibrational spectroscopy of anions has been performed by photoexcitation
spectroscopy into the neutral state using electron detachment 
\cite{kitsopoulos1990,xu1998}. Comparison of the resultant spectral features with
calculations reveals the structures of the anions. Recent work on larger
silicon cluster anions reveals vibrational spectra indicative of the
previously observed transition from prolate to spherical shapes \cite{meloni2004}.

The FELIX free electron laser source provides intense and tunable infrared radiation
including in the spectral range from 166 to 600 ~cm$^{-1}$ where silicon
clusters absorb. 
Vibrational spectroscopy of small silicon cluster cations was performed
using multiple photon dissociation spectroscopy. The ions were analysed
in a time-of-flight mass spectrometer. Also, isotopically selected
$^{129}$Xe atoms were attached to the clusters. Absorption of multiple
IR photons would lead to a depletion of the ion signal, allowing the
requisition of spectra of size-selected clusters.  
Comparison of the spectra with
calculations using density functional theory revealed novel structures
and a growth motif that started with a pentagonal bipyramid building block
and changed to a trigonal prism for larger clusters \cite{lyon2009}. 

Related work provided information on the structure of small neutral
silicon clusters \cite{fielicke2009}. Using a combination of tunable
far-infrared radiation from the FELIX free electron laser source and
vacuum-ultraviolet two-color ionisation, it was possible to scan the spectrum of
homonuclear neutral silicon clusters in the spectral range from 200 to
550 ~cm$^{-1}$. The use of vacuum-ultraviolet two-color ionisation
provided the advantage of detection of the initially neutral clusters
with mass selectivity. An increase 
of the ionisation rate was observed 
when IR photons had been absorbed. Comparison with density functional
theory (DFT) and M{\o}ller-Plesset (MP2) perturbation theory
calculations revealed that the ground state potential energy surface was
very flat. Therefore, rapid interconversion between different structures might well be
expected, as well as the presence of higher-energy isomers in
real-world experimental samples \cite{fielicke2009}.

Furthermore, theoretical work suggested that, at intermediate sizes,
around 20 atoms silicon clusters tend to build irregular cages
stabilised by a small number of encapsulated silicon atoms
\cite{mitas2000}. Silicon cages can also be stabilised by encapsulated
foreign atoms \cite{kumar2001,kumar2003,kumar2003prl}. Stable hollow
structures, similar to C$_{60}$ buckminsterfullerene (buckyballs), can
nevertheless be excluded \cite{sun2003}.    

In summary, the structures of silicon clusters grown in the
gas phase differ from the structures of silicon nanocrystals that have been
produced, for example, by etching of bulk crystalline silicon. The
structures of small silicon clusters are characterised by the tendency of the
atoms to minimise coordination, thereby favouring growth of prolate 
shapes. Therefore, their bond angles are smaller than their 
counterparts in sp$^3$ bonded, cubic diamond-structured bulk
silicon. The consequences of 
this behaviour are shorter internuclear separations and higher
atomic densities. An exemption from these principles is shown in the
work of Laguna and co-workers, however, who deposited silicon clusters produced
by pyrolysis of silane onto holey carbon films. High resolution
transmission electron images clearly shows nanoparticles with lattice
planes surrounded by an amorphous oxide shell \cite{hofmeister1999,laguna1999}.

\subsection[Electronic structure]{Electronic structure}

Photoelectron spectra of silicon cluster anions have frequently been
reported in the literature. These spectra exhibit rich features
for small clusters, which become smoother as the clusters become larger.
Maus, Gantef{\"o}r and Eberhardt \cite{maus2000} assign the low binding
energy features to the extra electron occupying the conduction bands.
The higher energy features observed are attributed to the valence
electrons. The energy difference between the two corresponds to the band
gap in the bulk picture. Small clusters between
$n$=3 and 13 were found to have band gap energies smaller than those typically seen for
 bulk crystals \cite{maus2000}. This is incommensurate with the idea of
quantum confinement, which would predict larger band gap energies for
anything smaller than the bulk. The
results were attributed to the entire geometric and electronic structure being affected
by surface effects, similar to the reconstruction of the surface of bulk
silicon crystals. While such an effect must clearly dominate the
electronic structure of small clusters, the trend continues for
larger clusters as well; Hoffmann et 
al., for instance, report the absence of a band gap for clusters up to 1000 atoms
\cite{hoffmann2001}. For large Si cluster anions, the photoelectron
spectrum is dominated by a single smooth and broad feature. The onset of
photoemission shifts with
size towards larger binding energies, a trend that is incommensurate
with a bulk band gap picture and contrary to what one would expect from
quantum confinement.

The observations made through photoelectron spectroscopy of cluster
anions in free beams match measurements of band gap energies of silicon
clusters deposited on HOPG \cite{marsen2000}. Such band gap energies were 
always smaller than those of their bulk counterparts, which agrees with a
different geometric structure than the diamond cubic structure of the
bulk. Also, a transition region was found for sizes around 1.5~nm,
corresponding to 44 atoms; larger clusters had no band gap. 

Band gap energies of cationic VSi$_n^+$ were determined by X-ray
spectroscopy using monochromatic synchrotron radiation and an ion trap
to store the size-selected clusters \cite{lau2011}. By measuring the ion
yield of specific ion decay channels, it was possible to record a direct 2p
photoionisation spectrum separately from the resonant 2p photoionisation
spectrum, yielding the energy difference $E_{XAS}$ between the core
level and the lowest unoccupied molecular orbital (LUMO,) and $E_{CL}$,
the energy difference between the core and the vacuum level,
$E_{VAC}$. Measurement of the valence state photoionisation spectrum $E_{VB}$
yields the difference between the highest occupied molecular orbital (HOMO)
and the the vacuum level, $E_{VAC}$. For the band gap energy, $E_{g}$, or, more
precisely, the HOMO-LUMO energy difference, it follows $E_{g} = E_{VB} -
E_{CL} + E_{XAS}$.

Photoionisation thresholds were measured for silicon clusters in the
size range from n~=~2 to 200 using laser photoionisation using
RETOF mass spectrometry detection \cite{fuke1993}. The ionisation
potential revealed features that were 
ascribed to a structural transition for sizes around n~=~20.
Measurements of the 2p core-level and valence electron binding energies
using monochromatic synchrotron radiation and an ion trap show a similar
size dependence \cite{vogel2012}. Both 2p binding energy and ionisation
potential show a 
linear dependence on the inverse cluster radius n$^{-1/3}$. Such a
dependence is expected from the size-dependent charging energy, similar
to metal clusters \cite{halder2015}. Furthermore, core level shifts were
derived and compared to ab initio calculations \cite{vogel2012}.

\section[Fluorescent silicon clusters]{Fluorescent silicon clusters}

The electronic level structure associated with dense packing,
suggesting very small band gap energies for small- and medium-sized
silicon clusters, and even the absence of band gaps, is unfavourable towards fluorescence. Indeed, 
fluorescence from free silicon clusters in traps or in molecular beams
has not yet been reported in the literature. The work on neutral and cationic silver
clusters in argon droplets and solid matrices
\cite{felix2001,sieber2004,conus2006,harbich2007} suggests that
fluorescence might be possible if silicon clusters are deposited and
embedded in rare gas matrices. While such deposition experiments have
been carried out \cite{honea1993,honea1999}, attempts to observe
fluorescence with this setup are not known to the author.

It appears that fluorescence reported for nanoscale silicon can be
attributed to the effects of quantum confinement, passivation,
and the presence of defects. To achieve confinement, passivation or
defects due to other materials, molecules or atoms are actively 
introduced. In the vast majority of bottom-up methods used to produce 
fluorescent silicon nanoparticles, chemical methods are
employed. Exceptions are laser vaporisation of silicon targets in
liquids \cite{svrcek2009,svrcek2009aging,intartaglia2012,alkis2012,svrcek2016,rodio2016}, pyrolysis of
silane in gas-flow reactors \cite{ehbrecht1997} and silicon cluster co-deposition with water
vapour \cite{haeften2009}.

\section[Red-orange luminescence]{Red-orange luminescence}

Silicon clusters produced by CO$_2$ laser-induced decomposition of
silane were found to show red photoluminescence after they were
deposited onto LiF or CaF$_2$ substrates and transferred to ambient air
\cite{ehbrecht1997}. Owing to a continuous supersonic beam and a pulsed
CO$_2$ laser, the part of the beam containing clusters was also pulsed. A
velocity selector was employed to select velocity segments of the
cluster pulse, and the cluster mass was measured by time-of-flight mass
spectrometry \cite{ledoux2002}. The average number of atoms in the
clusters, $\overline{N}$, 
was found to vary from $\overline{N}$~=~395, corresponding to an average diameter,
$\overline{D}$, of 2.47~nm to $\overline{N}$~=~9070, corresponding to
$\overline{D}$~=~7.03~nm. The diameters were deduced using a spherical
particle model, 
\begin{eqnarray}
D(N) & = & \left(\frac{3N}{4 \pi}V_{unit}\right)^{1/3} \label{eq:diameter}
\end{eqnarray}  
where V$_{unit}$~=~0.1601~nm$^3$ is the volume of the unit cell of bulk
crystalline silicon. Equation~\ref{eq:diameter} takes into account the fact that
the unit cell has a diamond cubic structure and contains eight atoms. It is
assumed that bulk and nanoparticle densities are the same.

These samples were photoexcited at 488~nm using continuous laser
radiation, and the resultant fluorescence spectrum was measured. Each sample showed
an almost symmetric fluorescent band whose peak wavelength decreased
with particle diameter. The size-dependent fluorescence wavelength
shifts agreed with the results one might anticipate from quantum confinement. Deviations were attributed to
partial oxidation of the surface layer, which could also be seen in high
resolution transmission electron microscopy (HRTEM)
images \cite{laguna1999}. The oxide shell thickness could be correlated linearly
to the cluster diameter. The smallest clusters of 6~nm in diameter
had an oxide shell with a thickness of 0.81~nm, whilst the largest particles had a
diameter of 34~nm had a 3~nm thick oxide shell \cite{hofmeister1999}.
The HRTEM images showed that the nanoparticles had a crystalline core.
A number of different silicon lattice planes were identified from
diffraction rings.

Hofmeister and co-workers also investigated the spacing of a (111)
lattice as a function of cluster size. They found that large silicon
clusters exhibited compressed (111) lattices compared to bulk
crystalline silicon. However, clusters smaller than 3~nm in diameter
were found to be dilated. The dependence of the (111) lattice spacing on
the cluster diameter followed:
\begin{eqnarray}
d(111) & = & \frac{0.023}{D} + 0.307 [nm]
\end{eqnarray}  
where $D$ is the cluster diameter.

Because the photoluminescence of silicon was found to depend on pressure,
the authors concluded that the size-dependent lattice separation must be
taken into account in a modified equation for the photoluminescence
energy caused by quantum confinement \cite{ledoux2000}.

\begin{eqnarray}
E_{PL}^{corr}  & = & E_{0} + \frac{3.73}{D^{1.39}} + \frac{0.881}{D} - 0.245
\end{eqnarray}  

Here, $E_{PL}^{corr}$ is the energy of the photoluminescent band, as
corrected for size-dependent lattice separation, in eV. $E_{0}$
is the band gap energy of bulk crystalline silicon at room temperature
(1.17 eV) \cite{ledoux2000}. 

It is important to note that all samples had been
transferred to air testing for photoluminescence. The samples were
kept in an argon atmosphere during 
transfer \cite{ehbrecht1997}. After production and exposure to air, the
crystalline core section of the particles was found to reduce in
diameter \cite{ledoux2000,ledoux2001}. 
Also, the photoluminescence energy was found to blue-shift, which was
attributed to the smaller sizes of the silicon clusters, supporting the
assertion that quantum confinement 
was controlling fluorescence properties \cite{ledoux2001}. The
effect of quantum confinement was also investigated by etching the oxide layer
using hydrofluoric acid (HF). This was found to narrow the spectral band
width of the luminescence but not the peak wavelength, since the silicon
crystalline core itself would clearly not be affected by such treatment
with HF. The effect of 
passivation on the fluorescence intensity, but not on the fluorescence
wavelength, is in line with earlier work by Seraphin and co-workers \cite{seraphin1996}.

Pyrolysis of silane in vacuum was also used by Li and co-workers to
produce silicon clusters \cite{li2004pyro,li2004langmuir}, who post-processed the
samples by etching with hydrofluoric acid (HF) and nitric acid (HNO$_3$).
This was found to reduce the cluster size and the intensity of visible luminescence.

\section[Blue luminescence]{Blue fluorescence}

\begin{figure}%[!ht] 
\centering
\includegraphics[width=0.9\columnwidth]{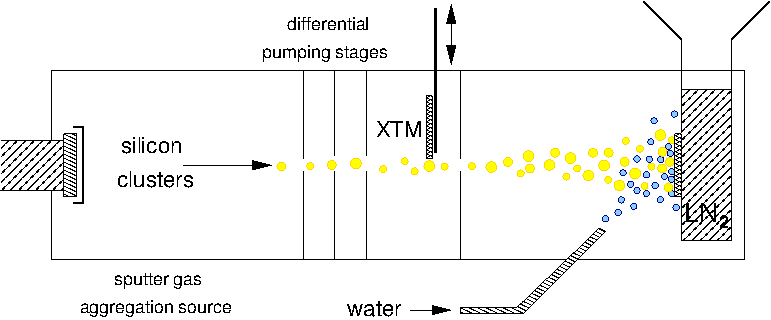}
\caption{
Schematic of experimental set-up used by \cite{haeften2009,brewer2009,torricelli2011}.
} 
\label{fig:sicluster-setup}
\end{figure}

To investigate the effect of passivation of silicon clusters in situ,
von Haeften and co-workers used a molecular beam co-deposition scheme
\cite{haeften2009}. They produced silicon clusters 
by gas aggregation using ion sputtering in an argon-helium atmosphere,
co-depositing them with a beam of water vapour onto a liquid
nitrogen-cooled target. After a deposition time of 30 minutes, the target was
warmed up, whereupon the ice-silicon mixture melted and a few millilitres of
liquid sample was collected. A schematic of the apparatus used is shown
in figure~\ref{fig:sicluster-setup}.

When photoexcited with 308~nm UV light, all liquid samples showed a blue
fluorescence that  
peaked at 420~nm \cite{haeften2009}. The fluorescence
intensity was stable over several months \cite{brewer2009}. 
When the photoexcitation wavelength was decreased from 310 to 240~nm, the
wavelength of the fluorescent band remained at 420~nm; however,
additional fluorescence bands appeared in the UV region
\cite{haeften2010,haeften2010sasp,torricelli2011}. When the clusters
were embedded in liquid ethanol and isopropanol, the fluorescence
wavelength shifted to 365 and 380~nm, respectively \cite{galinis2012bio}.
The fluorescence 
lifetime was measured using monochromatic synchrotron radiation. For an
excitation wavelength at 195~nm and fluorescence at 300~nm, a lifetime of
3.7~ns was determined \cite{yazdanfar2012}.

Time-correlated fluorescence spectroscopy showed that the blue
fluorescence consisted of at least two components: the spectral
component with a long fluorescent lifetime that peaked at 2.7~eV
\cite{yazdanfar2012}, which is a perfect 
match with the fluorescence of defect-rich silica
\cite{skuja1984}, and a second
short-lived, and much more intense, component, however, peaked at 3.0~eV
\cite{yazdanfar2012}. The good energy match and the long
lifetime of the first of these bands suggested it arose due to the spin-forbidden 
T$_1 \rightarrow $~S$_0$ transition observed for two-fold coordinated Si
in SiO$_2$ (O -- Si -- O)
\cite{skuja1984,skuja1992jns,nishikawa1992,fitting2004}. 
 
Short lifetimes have frequently been reported for the blue fluorescence
of nanoscale silicon \cite{kovalev1994}. 
Tsybeskov and co-workers investigated the lifetime of
blue fluorescence emitted from thermally and chemically oxidised porous
silicon \cite{tsybeskov1994}. The decay was multi-exponential with a
lifetime of $\sim$1~ns, which was independent of the excitation
energy. Furthermore, the appearance of blue fluorescence was correlated
to the presence of silicon oxide. Silicon-hydrogen bonds were absent. 

Harris and co-workers prepared blue light-emitting silicon samples by
electrochemical etching. They found that the samples rapidly
degraded, but were able to 
measure photoluminescence spectra at a sample temperature of 120~K;
both red- and blue-emitting components were observed. Fluorescence decay of the blue
fluorescence was measured at 77~K as having a time-constant of 0.86~ns. The
decay was mono-exponential \cite{harris1994}. 

By using different post-processing chemical treatments, porous silicon can be
prepared to emit either blue or red fluorescence. Zidek and co-workers used such
methods to measure the luminescence decay time separately for the two
different wavelength ranges. They found that both
types of fluorescence have distinct characteristics in their ultra-fast
decay time. The blue fluorescent band is attributed to several
underlying bands which vary further, depending on sample preparation \cite{zidek2011}.

Light emission in the blue spectral range is a known phenomenon for
colloidal suspensions of silicon nanoparticles
\cite{kimura1999,belomoin2000,valenta2008} and silicon nanocrystal films
\cite{loni1995,canham1996,deboer2010,ondic2014}. At the present time,
there seems to be a consensus 
that the vast majority of reported fluorescent bands in the blue spectral 
range are due to localised transitions, rather than being caused by quantum
confinement \cite{dasog2013,dasog2016}. The precise nature of this fluorescence is
nevertheless debated, though it is nevertheless
reasonable to attribute it to a range of different transitions which
have similar transition energies.

To elucidate the nature of the blue fluorescence, various sample
preparation techniques have been explored. Responses to various chemical
treatments and correlation to fluorescence performance, as well as chemical
analysis, have been employed. The results are not always consistent.

Konkievicz and co-workers prepared porous silicon films and investigated 
photoluminescence over a wide spectral range. Under photoexcitation with 193 nm excimer
laser radiation, 
they observed red luminescence. After annealing in an oxygen atmosphere
with 2\% organochlorine, blue luminescence around 400~nm appeared. The
blue band appeared at annealing temperatures of 750~$^0$C and increased
in intensity up to temperatures of 1050~$^0$C, after which no further
increases were seen \cite{kontkiewicz1994}. Fourier transform infrared
(FTIR) measurements showed the presence of silicon 
oxide. This preparation was also repeated in a nitrogen
atmosphere. Annealing in nitrogen did not produce blue
luminescence. This led to the conclusion that the blue luminescence
originates from oxidised nanostrutured silicon, although later work
showed a correlation between blue fluorescence intensity and nitrogen
content \cite{dasog2013}. Results similar to those of Konkievicz and co-workers   
have also been reported from porous Si that was oxidized and annealed at 880~$^0$C
\cite{kanemitsu1994}. Depending on the production method, the band
maxima range from 400 to 460~nm \cite{yu1998}.
The specific response of red and blue luminescence intensity to repeated
etching and oxidation was investigated by Lockwood and co-workers,
leading them to the conclusion that, at least, quantum 
confinement cannot be responsible for the blue luminescence
\cite{lockwood1996}.   

An important aspect of chemical treatment is how silicon nanoparticles
interact with an aqueous environment. This is because water is a strong
oxidising agent, but also because of the relevance of nanoparticles to
biomedical applications. Water may also be 
expected to quench fluorescence because of its dense vibrational
spectrum. The interaction with water has been found to chemically modify
silicon nanoparticles, and as a consequence its fluorescence activity has
been seen to 'degrade'
\cite{li2004water,erogbogbo2007,erogbogbo2008}. 
However, blue fluorescence from nanoscale silicon has been frequently observed, 
and specifically in connection with treatment with water
\cite{svrcek2009,svrcek2009aging,intartaglia2012ome,intartaglia2012,alkis2012,svrcek2016,rodio2016}.  

Hou and co-workers treated light-emitting porous silicon with boiling water
\cite{hou1993} and observed a large blue shift in the fluorescence towards the
green-blue spectral range. Infrared spectroscopy was performed with the
specific goal to assess
whether the formation of silicon monohydride (2080~cm$^{-1}$) and silicon
dihydride (2120~cm$^{-1}$) was correlated with the appearance of blue
fluorescence. Prior to treatment with water, both lines were
present in the spectrum, and decreased in intensity after boiling water
was added. Instead a band appeared at 1105~cm$^{-1}$, showing that
treatment with water had caused oxidation \cite{hou1993}.

Koyama and co-workers annealed oxidised porous silicon in water vapour
and observed a drastic enhancement in the blue fluorescence intensity
\cite{koyama1998}. Infrared
absorption spectroscopy indicated that this annealing increased
the absorption peaks related to OH vibrations except for those of free
silanol, which disappeared completely. No traces of carbon-related
signals were observed, contrary to the previously suspected involvement
of carbon \cite{kontkiewicz1994,canham1996}.

Many authors report that the emission of blue fluorescence is
correlated with very small structures, perhaps only a few nanometres in size. Akcakir and co-workers etched p-type boron-doped silicon using
H$_2$O$_2$ and HF \cite{yamani1997,akcakir2000}. The combined effect of the two
chemicals produced exceptionally small structures, which were then dispersed
in acetone. Under photoexcitation at 355~nm, blue fluorescence was
observed. Using two-photon excitation with 780~nm light pulses of 150~fs
duration, fluorescence correlation spectroscopy (FCS) was performed, which suggested
a hydrodynamic radius of 0.9~nm \cite{akcakir2000}. The same research group presented TEM
images of graphite films coated with this colloidal solution. The images
showed agglomerated particles of 1~nm in diameter, in very good
agreement with the FCS work \cite{belomoin2000}. IR spectra showed
various Si-H bands of the freshly prepared samples:
520-750~cm$^{-1}$ (SiH$_2$ scissors or SiH$_3$), 880-900~cm$^{-1}$ (Si-H
wagging) and 2070-2090~cm$^{-1}$ (SiH stretch and coupled H-Si-Si-H). The 
1070~cm$^{-1}$ Si-O stretch was also observed. Treatment with
H$_2$O$_2$ and subsequent IR spectroscopy was found to replace first the
di- and trihydrogen bonds, and then the Si-H with Si-O. The coupled
H-Si-Si-H bonds showed somewhat greater resilience. The blue
fluorescence intensity changed by no more than a factor of two after
H$_2$O$_2$ treatment \cite{belomoin2000}. 

Fluorescent silicon clusters produced by co-deposition with water vapour
onto a cold target showed a similar sizes. Atomic force microscopy
in non-contact and constant force mode of cluster films produced by
drop-casting colloidal solution onto freshly cleaved HOPG showed
uncovered regions of graphite, and agglomerated monolayers, as
well as double layers, of clusters \cite{torricelli2011}. The height of the monolayers
reflected the difference of the tip-HOPG and tip-cluster forces, and
hence cannot be taken as a measure of cluster height. However, measuring
the differences between the tip-first cluster and tip-second cluster
layers was expected to give a fair estimate of the height of the clusters in the
film. Values between 0.92 and 1.62~nm were found \cite{galinis2012surf}, in
very good agreement with Belomoin and co-workers \cite{belomoin2000}.  
 
Further studies using chemically produced silicon nanoparticles confirm
the relation between blue luminescence, small cluster sizes and
localised transitions. Zhong and co-workers report XRD diffraction peaks
similar to the   
diamond structure of bulk crystalline silicon \cite{zhong2013}.
Size distributions 
were measured by TEM, for which an average size of 2.2~nm was found \cite{zhong2013}.
Li and co-workers observed very high quantum yields of blue
fluorescence, up to 75\%, which 
were attributed to surface treatement with nitrogen-containing agents \cite{li2013}.  
Furthermore, Li showed that the fluorescence wavelength can be
tuned by different ligands attached to nitrogen-capped silicon
clusters \cite{li2016}. Also, the quantum yield could be increased
further, up to 90\% and the emission bandwidth could be narrowed. They
attribute their observations to localised transtions at 
the cluster surface and propose a `ligand-law' controlling the
photoluminescence \cite{li2016}.

\section[Conclusions]{Conclusions}

Silicon clusters consisting of a small number of atoms are fascinating 
objects through which one can study the evolution of material properties
with complexity and size. Free clusters produced in molecular beams have
properties that are unfavorable for light emission. However, when
passivated or embedded in a suitable host, they may emit 
fluorescence. The current available data show that both quantum
confinement and localised transitions, often at the surface,
are responsible for fluorescence. By building silicon clusters atom by atom, and by
embedding them in shells atom by atom, new insights into the microscopic
origins of fluorescence from nanoscale silicon can be expected.

The methods needed to perform such experiments, such as spectroscopy in droplets of
argon \cite{felix2001} and helium \cite{feng2015,katzy2016}, have
recently been developed. It can be hoped that they will be used for the
study of fluorescence of silicon clusters. In view of the rising number
of studies of fluorescent silicon nanostructures for biomedical
and other applications \cite{mcvey2014,dasog2014}, the value and
importance of such studies is clear.

\bibliographystyle{klausapalike}
%%%%%%%%%%%%%%add references to table of contents%%%%%%%%%%%%%
\addcontentsline{toc}{section}{References}
%%%%%%%%%%%%%%%%%%%%%%%%%%%%%%%%%%%%%%%%%%%%%%%%%%%%%%%%%%%%%%
%\bibliographystyle{agsm}
%\bibliographystyle{abbrvnat}
\setcitestyle{authoryear,open={(},close={)}}

\bibliography{abb_bibliography.bib}

\end{document}